\documentclass[aps,prl,groupedaddress, reprint,citeautoscript]{revtex4-1}

\usepackage{graphicx}
\usepackage{amsfonts}
\usepackage{amsmath}
\usepackage{amssymb}
\usepackage{bm}                    
\usepackage{dcolumn}               
\usepackage[multiple]{footmisc}    
\usepackage{setspace}              
\allowdisplaybreaks[3]  
\usepackage{color}

\newcommand{\Eq}    [1] {Eq.~(\ref{#1})}       
\newcommand{\Eqs}   [1] {Eqs.~(\ref{#1})}
\newcommand{\eq}    [1] {(\ref{#1})}
\newcommand{\Fig}   [1] {Fig.~\ref{#1}}

\newcommand{\eg}        {\textit{e.g.}}

\newcommand{\ie}        {\textit{i.e.}}

\newcommand{\rhs}       {r.h.s.}

\newcommand{\dyad}      {\otimes}

\newcommand{\hyd}         {\mathrm{H}}
\newcommand{\unspec}      {\circ}
\newcommand{\Trnspsd}     {\mathrm{T}}

\newcommand{\AngVel}      {\bm{\omega}}
\newcommand{\crtCoor}     {X}
\newcommand{\CrtCoor}     {\mathbf{\crtCoor}}
\newcommand{\crtVel}      {v}
\newcommand{\CrtVel}      {\mathbf{\crtVel}}

\newcommand{\rotMtrx}     {A}
\newcommand{\RotMtrx}     {\mathbf{\rotMtrx}}

\newcommand{\genCoor}     {Q}
\newcommand{\GenCoor}     {\mathbf{\genCoor}}

\newcommand{\genVel}      {U} 
\newcommand{\GenVel}      {\mathbf{\genVel}} 

\newcommand{\Epot}        {\Phi}
\newcommand{\force}       {f}
\newcommand{\Force}       {\mathbf{\force}}
\newcommand{\forge}       {F}
\newcommand{\Forge}       {\mathbf{\forge}}
\newcommand{\torq}        {\tau}
\newcommand{\Torq}        {\bm{\torq}}

\newcommand{\totFlow}     {u}
\newcommand{\TotFlow}     {\mathbf{\totFlow}}

\newcommand{\strain}      {E}
\newcommand{\Strain}      {\mathbf{\strain}}

\newcommand{\stress}      {S}
\newcommand{\Stress}      {\mathbf{\stress}}

\newcommand{\Mblty}       {\bm{\mu}}

\newcommand{\rsstnc}      {\xi}
\newcommand{\Rsstnc}      {\bm{\rsstnc}}
\newcommand{\rssqnc}      {\rho}
\newcommand{\Rssqnc}      {\bm{\rssqnc}}

\newcommand{\metric}      {g}

\newcommand{\viscosity}   {\eta}

\newcommand{\Random}      {\bm{\Theta}}

\newcommand{\GC}          {\GenCoor}

\newcommand{\GCt}      [1]{\GenCoor(#1)}

\newcommand{\GV}          {\GenVel}

\newcommand{\DGCt}     [1]{\Delta \GenCoor(#1)}

\newcommand{\RotMat}      {\RotMtrx }

\newcommand{\Avl}         {      \AngVel }

\newcommand{\Crt}         {\CrtCoor}

\newcommand{\Vel}         {      \CrtVel }

\newcommand{\DFrcBrwn}    {\delta \Force}

\newcommand{\DFrgBrwn}    {\delta \Forge}

\newcommand{\FrgHyd}      {      \Forge _\hyd}

\newcommand{\FrgPot}      {      \Forge _\Epot}
\newcommand{\Flt}         {      \TotFlow     _{\infty}}

\newcommand{\DTrqBrwn}    {\delta \Torq}

\newcommand{\Stn}         {      \Strain _{\infty}}

\newcommand{\MblUF}       {\Mblty^{\genVel   }_{\forge    }}

\newcommand{\MblUn}       {\Mblty^{\genVel   }_{\strain   }}

\newcommand{\MblsF}       {\Mblty^{\stress   }_{\forge    }}

\newcommand{\Mblsn}       {\Mblty^{\stress   }_{\strain   }}

\newcommand{\MblBB}       {      \Mblty                     }

\newcommand{\StsAv}       {\bar  {\Stress}     }

\newcommand{\Sts}         {       \Stress      }

\newcommand{\StsInt}      {       \Stress_\mathrm{I}}
\newcommand{\StsPot}      {       \Stress_\Epot}

\newcommand{\DSts}        {\Delta \Stress      }
\newcommand{\DStsBrwn}    {\delta \Stress      }
\newcommand{\DStsBrwnF}   {\delta \Stress_{\DFrgBrwn}}

\newcommand{\StsHyd}      {       \Stress _\hyd}

\newcommand{\Rst}         {\Rsstnc}
\newcommand{\RstFU}       {\Rsstnc^{\forge    }_{\genVel   }}

\newcommand{\RstFn}       {\Rsstnc^{\forge    }_{\strain   }}

\newcommand{\RstsU}       {\Rsstnc^{\stress   }_{\genVel   }}

\newcommand{\Rstsn}       {\Rsstnc^{\stress   }_{\strain   }}

\newcommand{\Rsq}         {\Rssqnc}

\newcommand{\RsqFU}       {\Rssqnc^{\forge    }_{\genVel   }}

\newcommand{\RsqFn}       {\Rssqnc^{\forge    }_{\strain   }}

\newcommand{\RsqsU}       {\Rssqnc^{\stress   }_{\genVel   }}

\newcommand{\Rsqsn}       {\Rssqnc^{\stress   }_{\strain   }}
\newcommand{\RsqsX}       {\Rssqnc^{\stress   }_{\unspec   }}

\newcommand{\mtrcGC}      {\metric_\genCoor}

\newcommand{\Dt}          {\Delta t}
\newcommand{\DtSqrt}      {\sqrt{\Dt}}
\newcommand{\einst}       {B}
\newcommand{\kB}          {k_B}

\newcommand{\viscSusp}    {\viscosity_s}
\newcommand{\viscFld}     {\viscosity_0}

\newcommand{\volCol}      {v} 
\newcommand{\volFrac}     {\phi}

\newcommand{\NablaGC}     {\Nabla_{\!\genCoor}}
\newcommand{\NablaCrt}    {\Nabla_{\!\crtCoor}}

\newcommand{\RndUt}       {\Random^\genVel(t)}
\newcommand{\Rndnt}       {\Random^\strain(t)}

\newcommand{\aspect}      {p}
\newcommand{\diameter}    {D}
\newcommand{\Dirac}       {\delta}

\newcommand{\length}      {L}

\newcommand{\Nabla}       {\bm{\nabla}}

\newcommand{\Zero}        {\mathbf{0}}

\newcommand{\GG}     {\mathbf{G}}

\begin{document}

\title{Fluctuating stresslets and the viscosity of colloidal suspensions}

\date{\today}


\author{Duraivelan Palanisamy}
\author{Wouter K. den Otter}
\email{w.k.denotter@utwente.nl}
\affiliation{
    Multi Scale Mechanics, Faculty of Engineering Technology
      and MESA+ Institute for Nanotechnology,\\
    University of Twente, P.O. Box 217, 7500 AE Enschede, The Netherlands
    }

\begin{abstract}
Theory and simulation of Brownian colloids
    suspended in an implicit solvent, 
      with the hydrodynamics of the fluid accounted for
        by effective interactions between the colloids,
  are shown to yield a marked and hitherto unobserved discrepancy between
  the viscosity calculated
    from the average shear stress under an imposed shear rate
      in the Stokesian regime
  and the viscosity extracted by the Green-Kubo formalism
    from the auto-correlations of thermal stress fluctuations
      in quiescent equilibrium.
We show that agreement between both methods is recovered
  by accounting for the fluctuating Brownian stresses on the colloids,
    complementing and related to
      the traditional fluctuating Brownian forces and torques
        through an extended fluctuation-dissipation theorem
          based on the hydrodynamic grand resistance matrix.
Time-averaging of the fluctuating terms
  gives rise to novel non-fluctuating stresslets.
Brownian Dynamics simulations of spheroidal particles
  illustrate the necessity of these
    fluctuating and non-fluctuating contributions 
      to obtaining consistent viscosities.
\end{abstract}

\maketitle


\section{\label{intro}Introduction}

Einstein derived in his thesis
  that adding rigid spherical colloids
    to a Newtonian fluid of viscosity $\viscFld$
  creates a suspension of effective viscosity
\begin{equation}
  \label{Einstein}
    \viscSusp
  =
    \viscFld
    \left(
      1 + \einst \volFrac
    \right),
\end{equation}
  with Einstein coefficient $\einst = 5/2$,
    for low colloidal volume fractions $\volFrac$ 
  \cite{Einstein1906, Einstein1911}.
This celebrated result,
    based on the analytic solution of
      Stokesian straining flow around a spherical particle
      \cite{KimKarrila,GuazzelliMorris},
  is readily reproduced by Brownian Dynamics (BD) simulations
    of an isolated spherical colloid suspended in a fluid
      subject to a linear shear flow.
Viscosities can also be determined from quiescent fluids,
  using the Green-Kubo formalism
    of integrating the auto-correlations
      of the spontaneous stress fluctuations
      \cite{McQuarrie}.
Rather surprisingly,
  both the aforementioned theory and BD simulations
      of an isolated spherical particle
    then yield $\einst = 0$,
      as will be demonstrated below.
One might argue that this difference
    is an artefact of studying a one-particle system,
  which could be the reason that it appears
    not to have been discussed before in the literature,
  but we are of the opinion that
    it reveals a deficiency in the current understanding
      of stress calculations of suspensions of Brownian particles.
We propose a solution,
    the inclusion of fluctuating Brownian stresses,
  that recovers agreement
      between equilibrium and non-equilibrium evaluations
        of the Einstein coefficient of a spherical particle,
          at $\einst = 5/2$.
By expressing the stochastic equations
    for the motion and stress in the It\^o form,
  two novel non-vanishing stress contributions emerge
    from correlations between the various fluctuations.
These terms affect
  the Einstein coefficients of isolated non-spherical particles,
    both in quiescent fluids and in flowing fluids.
The novel terms also contribute to
    the viscosities of non-dilute solutions,
  including suspensions of spherical particles
    \cite{Batchelor72,Batchelor77}.


\section{\label{sec:ConvBD}Brownian motion}

Consider a non-Brownian particle with generalized velocity
  $\GV = ( \Vel, \Avl )$,
    where $\Vel$ and $\Avl$ denote linear and angular velocity,
      respectively,
  in an incompressible fluid subject to a linear flow field
    characterized by the local generalized velocity $\Flt$
      and the strain rate tensor $\Stn$,
        {\ie} the traceless symmetric
          $(3 \times 3)$ velocity gradient matrix.
In the limit of Stokesian flow,
  the generalized hydrodynamic force $\FrgHyd$,
      a vector comprizing
        three force components and three torque components,
    and the deviatoric hydrodynamic stress $\StsHyd$,
        a traceless symmetric $(3 \times 3)$ matrix,
      acting on the particle are given by
        \cite{Durlofsky87,KimKarrila,GuazzelliMorris}
\begin{equation}
  \label{ResistancePicture}
    \left(
      \begin{array}{c}
        \FrgHyd \\
        \StsHyd
      \end{array}
    \right)
  =
  - \left(
      \begin{array}{cc}
        \RstFU & \RstFn \\
        \RstsU & \Rstsn
      \end{array}
    \right)
    \left(
      \begin{array}{r@{\;-\;}l}
        \GV & \Flt \\
            & \Stn
      \end{array}
    \right),
\end{equation}
  where $\Rst$ is the grand resistance matrix
    \cite{KimKarrila,GuazzelliMorris},
  whose four parts are labeled with
    a  lower index specifying the multiplication partner and
    an upper index highlighting the ensuing result.
Upon neglecting inertial effects,
  the equation of motion for a colloid
    experiencing also a generalized potential-derived force $\FrgPot$
      and a fluctuating Brownian force $\DFrgBrwn$
  is solved from a balance of forces,
    $\FrgPot + \FrgHyd + \DFrgBrwn = \Zero$.
A partial inversion of the above equation then gives
        \cite{Durlofsky87,KimKarrila,GuazzelliMorris}
\begin{equation}
 \label{EqOfMotionOld}
    \left(
      \begin{array}{c}
        \GV \\
        \Sts
      \end{array}
    \right)
  =
    \left(
      \begin{array}{cc}
        \MblUF & \MblUn \\
        \MblsF & \Mblsn
      \end{array}
    \right)
    \left(
      \begin{array}{c}
        \FrgPot + \DFrgBrwn \\
      - \Stn
      \end{array}
    \right)
  +
    \left(
      \begin{array}{c}
        \Flt  \\
        \Zero
      \end{array}
    \right),
\end{equation}
  where $\MblBB$ is the generalized mobility matrix,
      again expressed as a combination of four labeled parts,
    and $\Sts = - \StsHyd$ denotes the stress exerted on the fluid
      by the moving colloid.
It is important to realize that
  the stress $\Sts$ is \textit{not} the result of a balance of stresses,
    but a direct consequence of
      the velocity difference between colloid and fluid.
Should one desire so,
    for instance when studying easily deformable particles,
  a balance must be constructed between
    the total stress acting on the colloid
      --~due to
        the hydrodynamic stress $\StsHyd$,
        a potential-derived $\StsPot$
        and Brownian contributions (as discussed below)~--
    and the internal elastic stress of the particle $\StsInt$;
  after solving this balance for the unknown $\StsInt$,
    the combination thereof with the elasticity tensor of the colloid
      yields its deformation.
We will here consider rigid particles instead,
  and note for completeness that
    their stress balances are closed
      by unspecified Lagrange multipliers for $\StsInt$
        at vanishing deformation.
The random force perturbations $\DFrgBrwn$
  have zero mean,
  are uncorrelated in time (Markovian)
  and obey the classical fluctuation-dissipation theorem
    derived from the symmetric positive-definite $(6 \times 6)$
      force-velocity segment of the resistance matrix
        \cite{Ermak78,Brady93},
\begin{equation}
  \label{FlucDissOld}
    \langle
      \DFrgBrwn( t  )
      \dyad
      \DFrgBrwn( t' ) 
    \rangle
  =
    2 \kB T
    \RstFU
    \Dirac( t - t' ),
\end{equation}
  where $t$ and $t'$ denote times,
    $\kB$ Boltzmann's constant,
    $T$ the temperature
    and $\Dirac$ the Dirac delta function.
The textbook proof of the fluctuation-dissipation theorem
  is its ability,
    in combination with a second-order Langevin equation of motion,
  to reproduce the Maxwell-Boltzmann equilibrium velocity distribution
    \cite{McQuarrie,Kampen}.
Following the introduction of fluctuating hydrodynamics
    by Landau and Lifshitz \cite{Landau57,LifshitzPitaevskii},
  several authors have shown that the above theorem for a colloid
    also follows from the fluctuation-dissipation theorem of the fluid
      \cite{Zwanzig64,Fox70,Hauge73,Bedeaux74,Noetinger90,Singh17}.
Note that
  the force perturbations
    affect the velocity difference between colloid and fluid
  and thereby give rise to
    an indirect Brownian contribution to the stress \cite{Bossis89},
    $\DStsBrwnF = \MblsF \DFrgBrwn(t)$,
 as follows from \Eq{EqOfMotionOld}.
   
For a free spherical particle of volume $\volCol$ in a linear shear flow,
  the above expressions give rise to the average stress
    $\langle \Sts \rangle = - \Mblsn \Stn$,
    where the minus sign indicates resistance to the flow.
Inserting the theoretical expression for $\Mblsn$
      \cite{KimKarrila,GuazzelliMorris}
  then yields $\einst = 5/2$, as expected.
But applying the Green-Kubo formalism
    to the spontaneous stress fluctuations
      $\DSts(t) = \Sts    (t) - \langle \Sts     \rangle$
      in a quiescent fluid \cite{McQuarrie,Daivis94},
\begin{align}
  \label{GreenKubo}
    \einst
  =
    \frac{
      1
    }{
      10
      \kB T
      \viscFld
      \volCol
    }
    \int_0^\infty
      \langle \DSts(t) : \DSts(0) \rangle
    \, d t,
\end{align}
  yields $\einst = 0$ for a spherical particle,
    as follows from observing that under these conditions
      $\DSts(t) = \DStsBrwnF(t) = \Zero$
      since $\MblsF = \Zero$ for a spherical particle.
It is this disconcerting discrepancy between the $\einst$-s
  that motivates the current research.


\section{Fluctuating Brownian stress}

A suspended colloidal particle experiences a myriad of collisions
  due to the thermal motions of the surrounding solvent molecules.
The sum over all collisions over a short time interval
  --~sufficiently short to ignore the motion of the colloid,
     yet encompassing a large number of molecular collisions~--
     gives rise to the fluctuating Brownian force $\DFrgBrwn(t)$
       discussed above.
Note that this generalized force
    comprises a force $\DFrcBrwn(t)$ and a torque $\DTrqBrwn(t)$,
  which represent distinct `projections' of the same molecular noise
    integrated over the surface of the colloid
      \cite{Batchelor70,Durlofsky87,GuazzelliMorris}.
Their common origin implies that the force and torque are correlated,
  as reflected by the fluctuation-dissipation theorem
    in \Eq{FlucDissOld}.
It is only natural to assume
  that the collisions also give rise
    to a fluctuating Brownian stress on the particle, $\DStsBrwn$,
    {\ie} a stress distribution over the surface
      that would cause a soft particle to deform,
  which constitutes a third projection of the same molecular noise.
This direct fluctuating Brownian stress $\DStsBrwn$
    is not to be confused with $\DStsBrwnF$,
  the latter being an indirect fluctuating stress
    resulting from a velocity difference between colloid and fluid
      caused by the first and second projections of the thermal noise.
We next need to determine the strength
    of the fluctuating Brownian stress $\DStsBrwn$,
  which in view of the preceding discussion does not follow
    from a stress balance on the particle.
To conform with common practice in the field
    \cite{KimKarrila,GuazzelliMorris},
  our interest here will be on the deviatoric parts of the stresses.

For any (non-Brownian) colloid experiencing a flow field,
  the hydrodynamic force acting on the particle is obtained as
    the zeroth moment of the traction vector integral over the surface
  while the torque and the stress or `stresslet' on the particle
    are given by
      (a permutation of) the anti-symmetric and the symmetric
        first moments of the traction vector integral over the surface,
        respectively
      \cite{Batchelor70,Durlofsky87,GuazzelliMorris}.
If this flow field is replaced
    by the fluctuating hydrodynamics of the fluid,
  the strengths of the resulting fluctuating Brownian force $\DFrcBrwn$
    and torque $\DTrqBrwn$,
      as well as their cross-correlations,
    are given by the fluctuation dissipation theorem
      of \Eq{FlucDissOld}.
We now hypothesize that
  the fluctuations of $\DFrgBrwn$ and $\DStsBrwn$,
      given their common origin
        as projections of fluctuating hydrodynamics,
    are related by a generalized fluctuation-dissipation theorem
      based on the grand resistance matrix,
\begin{equation}
  \label{FlucDissNew}
    \left\langle
      \left(
        \begin{array}{c}
          \DFrgBrwn( t  ) \\
          \DStsBrwn( t  )
        \end{array}
      \right)
      \dyad
      \left(
        \begin{array}{c}
          \DFrgBrwn( t' ) \\
          \DStsBrwn( t' )
        \end{array}
      \right) 
    \right\rangle
  =
    2 \kB T
    \left(
      \begin{array}{cc}
        \RstFU & \RstFn \\
        \RstsU & \Rstsn
      \end{array}
    \right)
    \Dirac( t - t' ).
\end{equation}
Whereas the force correlations in this expression are well established
      \cite{Zwanzig64,Fox70,Hauge73,Bedeaux74,Noetinger90,Singh17},
  the stress correlations and the force-stress cross-correlations
    have attracted little attention.
A recent study supports the validity of our assumption
    for spherical particles,
  however without providing an explicit expression \cite{Singh17}.
Introducing the fluctuating stresslet into \Eq{EqOfMotionOld} gives
  the extended expression for the motion and the stress,
\begin{equation}
 \label{EqOfMotionNew}
    \left(
      \begin{array}{c}
        \GV \\
        \Sts
      \end{array}
    \right)
  =
    \left(
      \begin{array}{cc}
        \MblUF & \MblUn \\
        \MblsF & \Mblsn
      \end{array}
    \right)
    \left(
      \begin{array}{c}
        \FrgPot + \DFrgBrwn \\
      - \Stn
      \end{array}
    \right)
  +
    \left(
      \begin{array}{c}
        \Flt \\  
        \DStsBrwn
      \end{array}
    \right).
\end{equation}
It is evident that this expression produces
  identical translational and rotational Brownian motion
    to the classical expression,
  even though the fluctuating force is now correlated
    to the fluctuating stresslet.
With the inclusion of the fluctuating stresslet,
  the analytic calculations of the viscosity
    for an isolated force-free rigid spherical particle
      under equilibrium and non-equilibrium conditions
  are now in agreement,
    both yielding $\einst = 5/2$
      (and this result is unaffected by the full derivation below).

The interpretation of
  the stochastic differential equation of motion, \Eq{EqOfMotionNew},
    requires further attention to resolve an ambiguity:
  the impact of the Brownian force,
    as determined by
          the resistance matrix in the fluctuation-dissipation theorem
          and the mobility matrix in the equation of motion,
    varies with the coordinates $\GC$ of the particle,
  while the coordinates change due to this Brownian force.
In the It\^o interpretation,
    {\ie} using only parameter values at time $t$
      just before the impact of the Brownian force,
  the integration of the equation of motion over a time step $\Dt$
    results in coordinate increments \cite{Palanisamy19}
\begin{align}
  \label{EqOfMotionBD}
  \begin{split}
    \DGCt{t}
 &=
    \GCt{ t + \Dt } - \GCt{t}
\\
 &=
  - \MblUF
    \NablaGC
      \Big(
        \Epot - \ln \mtrcGC^{1/2}
      \Big)
    \Dt
  -
    \MblUn
    \Stn
    \Dt
\\
 &\phantom{==}
  +
    \Flt
    \Dt
  +
    \kB T
    \NablaGC
    \cdot
    \MblUF
    \Dt
\\
 &\phantom{==}
  +
    \MblUF
    \left[
      \RsqFU
      \RndUt
    +
      \RsqFn
      \Rndnt
    \right]
    \DtSqrt,
  \end{split}
\end{align}
  where $\mtrcGC$ denotes the metric of the coordinate space,
    $\Rsq$ is the symmetric tensor solving $\Rsq^2 = 2 \pi \kB T \Rst$,
  and $\RndUt$ and $\Rndnt$ are random vectors
      with zero mean, unit variance and devoid of correlations,
    containing six and five unique elements, respectively.
The divergence term brings into account
    the coordinate-dependence of the hydrodynamic matrices
       \cite{Ermak78,Kampen,Ottinger,Gardiner};
  these additional displacements are not evident from \Eq{EqOfMotionNew}
    but are crucial
        to obtaining the equilibrium Boltzmann distribution
      and, as will be shown below,
        also contribute to the stress.
An alternative interpretation, due to Einstein,
  gave rise to the name `thermodynamic force'
    \cite{Einstein05,Batchelor76,GuazzelliMorris}.

On a technical note,
  since in the current context
    the stress and strain rate tensors
      are symmetric and traceless $( 3 \times 3)$ matrices,
  it proves convenient to replace both by five-vectors
    so the usual mathematical and numerical techniques
      can be applied to the resulting symmetric
        $(11 \times 11)$ hydrodynamic matrices
          \cite{KimKarrila,Palanisamy18}.
Because the hydrodynamic matrices and the conservative potential
  are typically expressed in terms of
    Cartesian velocities and Cartesian forces,
      and in angular velocities and torques around Cartesian axes,
  we furthermore take the freedom
    of evaluating the {\rhs} of \Eqs{EqOfMotionNew} and \eq{EqOfMotionBD}
      in Cartesian coordinates,
    henceforth collectively denoted as $\Crt$.
Since the Cartesian angular velocities
    are not time derivatives of angular coordinates,
  the rotation angle increments still require transformation
      to proper generalized coordinates $\GC$
    describing the orientation of the colloid
      in terms of {\eg} Euler angles or quaternions,
    or one may directly update the rotation matrix
      between the colloid-based axes frame and the space-based axes frame.
The latter two options have the advantage
  that they do not require corrections resulting from the metric.
    \cite{Makino04motion,Ilie15,Palanisamy18}.

Continuing in the It\^o representation,
  we find by some mathematical manipulations
    of \Eqs{FlucDissNew} through \Eq{EqOfMotionBD}
  that the average stress exerted by the colloid on the fluid
    during a time step $\Dt$ reads as \cite{Palanisamy19}
\begin{align}
  \label{StressNew}
  \begin{split}
    \StsAv(t)
 &=
    \frac{ 1 }{ \Dt }
    \int_t^{ t + \Dt }
      \Sts(t')
    d t'
\\
 &=
    \MblsF
    \FrgPot
  -
    \Mblsn
    \Stn
  +
    \NablaCrt
    \cdot
    \MblsF
\\
 &\phantom{==}
  +
    \left[
      \left(
        \NablaCrt
          \RsqsU
      \right)
      \RsqFU
    +
      \left(
        \NablaCrt
          \Rsqsn
      \right)
      \RsqFn
    \right]
    :
    \MblUF
\\
 &\phantom{==}
  +
    \frac{ 1 }{ \DtSqrt }
    \Big\{
      \MblsF
      \left[
        \RsqFU
        \RndUt
      +
        \RsqFn
        \Rndnt
      \right]
\\
 &\phantom{======}
    +
      \RsqsU
      \RndUt
    +
      \Rsqsn
      \Rndnt
    \Big\}.
  \end{split}
\end{align}
The first and second term on the {\rhs} are
  the two deterministic contributions to \Eq{EqOfMotionNew}.
The third term,
  usually derived along another route,
    results here by combining
      a term related to the divergence in \Eq{EqOfMotionBD}
      with a term resulting from
        correlations between $\DFrgBrwn$ and $\DStsBrwnF$,
  and is referred to in the literature as
    `Brownian stress' or `diffusion stress'
    \cite{Batchelor77, Rallison78, Bossis89, Makino04visco}.
The fluctuating terms,
    collected between curly brackets in the last term to \Eq{StressNew},
  have zero average
    and may therefore be ignored
      when calculating the time-averaged stress
        of a system under shear flow,
  but their correlations are crucial
    when applying the Green-Kubo formalism to a quiescent system.
In both non-equilibrium and equilibrium cases,
  the time-averaged correlations
    of the fluctuating forces and the fluctuating stresslets
      give rise to two additional non-fluctuating stress contributions,
    the two $\NablaCrt \RsqsX$ terms in \Eq{StressNew},
      which, to the best of our knowledge,
        are derived and reported here for the first time.


\section{Numerical example}

\begin{figure} 
  \centering
    \includegraphics[width=7.2cm]{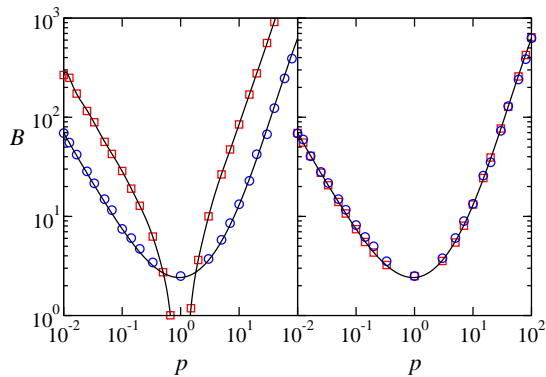}
  \caption{
    \label{fig:spheroids}
    Einstein viscosity coefficients $\einst$
      for isolated rigid spheroidal particles
        of various aspect ratio $\aspect$,
      as deduced from
        the average stress in simulations at low shear rate (blue circles)
        and the stress fluctuations at zero shear (red squares).
    Simulations
      based on the classical stress calculation of \Eq{EqOfMotionOld}
        show a marked difference between the two approaches (left),
      while the revisions proposed in \Eq{EqOfMotionNew}
        recover agreement (right).
    The smooth lines are guides to the eye.
  }
\end{figure}

As an illustration of the revised stress calculation,
  we present numerical simulations of isolated spheroidal particles.
For a rigid particle,
  the hydrodynamic matrices are constant in the body frame
    and rotate with the body in the space frame,
  which permits efficient calculations of the motion and the stress
    without demanding re-evaluation of the hydrodynamics
      \cite{Makino04motion,Ilie15,Palanisamy18}.
With $\RotMat$ denoting the body-to-space rotation matrix,
  the three derivative-containing terms in \Eq{StressNew}
    turn out to be of the form $\RotMat \GG \RotMat^\Trnspsd$,
      where $\GG$ is a body-dependent constant $(3 \times 3)$ matrix
        and the superscript $\Trnspsd$ indicates transposition.
Simulations of a spherical particle
    confirm the theoretical results mentioned above,
  with the classical approach of \Eq{EqOfMotionOld}
    yielding Einstein coefficients of 5/2 under shear
      and zero by the Green-Kubo method,
  while the amendments proposed here yield $\einst = 5/2$
    for both non-equilibrium and equilibrium simulations.
Numerical results for the Einstein coefficients of spheroidal particles
  are presented in \Fig{fig:spheroids}
    as function of the aspect ratio $\aspect = \length / \diameter$
      between the length $\length$ along the symmetry axis
        and the diameter $\diameter$ in the perpendicular direction,
    ranging from disk-like oblate to needle-like prolate.
For simulations based on \Eq{EqOfMotionOld},
  applying the Green-Kubo formalism to quiescent systems
    yields Einstein coefficients
      between four ($\aspect \ll 1$) and eight ($\aspect \gg 1$)
        times higher than the values obtained
          from the average stress in sheared systems,
        with an intermediate dip of $\einst$ approaching zero
          for near-spherical particles.
Simulations based on \Eq{EqOfMotionNew}, however,
  show consistency
    between equilibrium and non-equilibrium viscosity evaluations,
  see \Fig{fig:spheroids}.
The biggest difference between the classical and the proposed methods
  is in the Einstein coefficients
    deduced from the thermal stress fluctuations in quiescent fluids,
  and mainly results from
    the inclusion of the Brownian fluctuating stress in \Eq{StressNew}.
The two novel $\NablaCrt \RsqsX$ terms in said expression
  introduce a relatively modest increase
    of the Einstein coefficient obtained under shear,
  by about 1\% at $\aspect = 20$ and 5\% at $\aspect = 100$.
The impact of these terms is larger for less symmetric bodies,
  amounting to
    about $+10\%$ for a semi-disk
      with diameter-to-thickness ratio of 40
  and about $-8\%$ for a helix inscribing 7.5 revolutions
    in a cylinder with a length-to-diameter ratio of five.


\section{Conclusions}

The perpetual thermal motion of fluids
    contributes to the viscosity of colloidal suspensions,
  both 
    by causing the Brownian motion of the colloids
      and by inducing flucuating stresses on the colloids.
Inclusions of these Brownian stresses,
    absent in current theoretical and numerical
      implicit solvent methods for suspensions,
  is therefore necessary to obtain the correct viscosity.
Our theoretical analysis of dilute suspensions of spherical particles
    and numerical simulations of spheroids 
  illustrate the validity of the amendments proposed
    in the Brownian motion and stress calculation of \Eq{EqOfMotionNew}
    and
    the extended fluctuation-dissipation theorem of \Eq{FlucDissNew}.
Correlations between the various fluctuating terms
  then give rise to novel non-fluctuating contributions
      in the It\^o representation of the stress, see \Eq{StressNew}.
A more detailed exposition of the derivations outlined above,
    along with additional numerical results on colloids of various shapes,
  will be presented elsewhere \cite{Palanisamy19}.
The expression for the time-averaged stresslet on a colloid
    is readily extended to a collection of $N$ particles,
  by enlarging the mobility and resistance matrices
    to $(11 N \times 11 N)$ matrices
      including hydrodynamic interactions between all colloids;
  the total deviatoric stress in the system is then obtained
    by adding up
      the stresslets of the individual particles,
      the virial term due to
        generalized conservative forces on the colloids, and
      the shear resistance of the suspending fluid.


\section*{Acknowledgments}

We thank Prof.~Stefan Luding for stimulating discussions.
This work is part of
  the Computational Sciences for Energy Research
    Industrial Partnership Programme
  co-financed by Shell Global Solutions B.V.
  and the Netherlands Organisation for Scientific Research (NWO).



\begin{thebibliography}{10}%
\makeatletter
\providecommand \@ifxundefined [1]{%
 \ifx #1\undefined \expandafter \@firstoftwo
 \else \expandafter \@secondoftwo
\fi
}%
\providecommand \@ifnum [1]{%
 \ifnum #1\expandafter \@firstoftwo
 \else \expandafter \@secondoftwo
\fi
}%
\providecommand \enquote [1]{``#1''}%
\providecommand \bibnamefont  [1]{#1}%
\providecommand \bibfnamefont [1]{#1}%
\providecommand \citenamefont [1]{#1}%
\providecommand\href[0]{\@sanitize\@href}%
\providecommand\@href[1]{\endgroup\@@startlink{#1}\endgroup\@@href}%
\providecommand\@@href[1]{#1\@@endlink}%
\providecommand \@sanitize [0]{\begingroup\catcode`\&12\catcode`\#12\relax}%
\@ifxundefined \pdfoutput {\@firstoftwo}{%
 \@ifnum{\z@=\pdfoutput}{\@firstoftwo}{\@secondoftwo}%
}{%
 \providecommand\@@startlink[1]{\leavevmode}%
 \providecommand\@@endlink[0]{}%
}{%
 \providecommand\@@startlink[1]{%
  \leavevmode
  \pdfstartlink
   attr{/Border[0 0 1 ]/H/I/C[0 1 1]}%
   user{/Subtype/Link/A<</Type/Action/S/URI/URI(#1)>>}%
  \relax
 }%
 \providecommand\@@endlink[0]{\pdfendlink}%
}%
\providecommand \url  [0]{\begingroup\@sanitize \@url }%
\providecommand \@url [1]{\endgroup\@href {#1}{\urlprefix}}%
\providecommand \urlprefix [0]{URL }%
\providecommand \Eprint[0]{\href }%
\@ifxundefined \urlstyle {%
  \providecommand \doi [1]{doi:\discretionary{}{}{}#1}%
}{%
  \providecommand \doi [0]{doi:\discretionary{}{}{}\begingroup
  \urlstyle{rm}\Url }%
}%
\providecommand \doibase [0]{http://dx.doi.org/}%
\providecommand \Doi[1]{\href{\doibase#1}}%
\providecommand \bibAnnote [3]{%
  \BibitemShut{#1}%
  \begin{quotation}\noindent
    \textsc{Key:}\ #2\\\textsc{Annotation:}\ #3%
  \end{quotation}%
}%
\providecommand \bibAnnoteFile [2]{%
  \IfFileExists{#2}{\bibAnnote {#1} {#2} {\input{#2}}}{}%
}%
\providecommand \typeout [0]{\immediate \write \m@ne }%
\providecommand \selectlanguage [0]{\@gobble}%
\providecommand \bibinfo [0]{\@secondoftwo}%
\providecommand \bibfield [0]{\@secondoftwo}%
\providecommand \translation [1]{[#1]}%
\providecommand \BibitemOpen[0]{}%
\providecommand \bibitemStop [0]{}%
\providecommand \bibitemNoStop [0]{.\EOS\space}%
\providecommand \EOS [0]{\spacefactor3000\relax}%
\providecommand \BibitemShut [1]{\csname bibitem#1\endcsname}%
\bibitem{Einstein1906}%
  \BibitemOpen
  \bibfield{author}{%
  \bibinfo {author} {\bibfnamefont{A.}~\bibnamefont{Einstein}},\ }%
  \bibfield{journal}{%
  \bibinfo {journal} {Ann. Phys.}\ }%
  \textbf{\bibinfo {volume} {19}},\ \bibinfo {pages} {289} (\bibinfo {year}
  {1906})%
  \bibAnnoteFile{NoStop}{Einstein1906}%
\bibitem{Einstein1911}%
  \BibitemOpen
  \bibfield{author}{%
  \bibinfo {author} {\bibfnamefont{A.}~\bibnamefont{Einstein}},\ }%
  \bibfield{journal}{%
  \bibinfo {journal} {Ann. Phys.}\ }%
  \textbf{\bibinfo {volume} {34}},\ \bibinfo {pages} {591} (\bibinfo {year}
  {1911})%
  \bibAnnoteFile{NoStop}{Einstein1911}%
\bibitem{KimKarrila}%
  \BibitemOpen
  \bibfield{author}{%
  \bibinfo {author} {\bibfnamefont{S.}~\bibnamefont{Kim}}\ and\ \bibinfo
  {author} {\bibfnamefont{S.~J.}\ \bibnamefont{Karrila}},\ }%
  \emph{\bibinfo {title} {Microhydrodynamics: Principles and Selected
  Applications}},\ Butterworth-Heinemann Series in Chemical Engineering\
  (\bibinfo {publisher} {Butterworth-Heinemann},\ \bibinfo {address} {Stoneham,
  MA, U.S.A.},\ \bibinfo {year} {1991})%
  \bibAnnoteFile{NoStop}{KimKarrila}%
\bibitem{GuazzelliMorris}%
  \BibitemOpen
  \bibfield{author}{%
  \bibinfo {author} {\bibfnamefont{E.}~\bibnamefont{Guazzelli}}\ and\ \bibinfo
  {author} {\bibfnamefont{J.~F.}\ \bibnamefont{Morris}},\ }%
  \emph{\bibinfo {title} {A Physical Introduction to Suspension Dynamics}},\
  Cambridge Texts in Applied Mathematics\ (\bibinfo {publisher} {Cambridge
  University Press},\ \bibinfo {address} {Cambridge, U.K.},\ \bibinfo {year}
  {2012})%
  \bibAnnoteFile{NoStop}{GuazzelliMorris}%
\bibitem{McQuarrie}%
  \BibitemOpen
  \bibfield{author}{%
  \bibinfo {author} {\bibfnamefont{D.~A.}\ \bibnamefont{McQuarrie}},\ }%
  \emph{\bibinfo {title} {Statistical {M}echanics}}\ (\bibinfo {publisher}
  {Harper \& Row Publishers},\ \bibinfo {address} {New York, NY, U.S.A.},\
  \bibinfo {year} {1976})%
  \bibAnnoteFile{NoStop}{McQuarrie}%
\bibitem{Batchelor72}%
  \BibitemOpen
  \bibfield{author}{%
  \bibinfo {author} {\bibfnamefont{G.~K.}\ \bibnamefont{Batchelor}}\ and\
  \bibinfo {author} {\bibfnamefont{J.~T.}\ \bibnamefont{Green}},\ }%
  \bibfield{journal}{%
  \bibinfo {journal} {J. Fluid Mech.}\ }%
  \textbf{\bibinfo {volume} {56}},\ \bibinfo {pages} {401} (\bibinfo {year}
  {1972})%
  \bibAnnoteFile{NoStop}{Batchelor72}%
\bibitem{Batchelor77}%
  \BibitemOpen
  \bibfield{author}{%
  \bibinfo {author} {\bibfnamefont{G.~K.}\ \bibnamefont{Batchelor}},\ }%
  \bibfield{journal}{%
  \bibinfo {journal} {J. Fluid Mech.}\ }%
  \textbf{\bibinfo {volume} {83}},\ \bibinfo {pages} {97} (\bibinfo {year}
  {1977})%
  \bibAnnoteFile{NoStop}{Batchelor77}%
\bibitem{Durlofsky87}%
  \BibitemOpen
  \bibfield{author}{%
  \bibinfo {author} {\bibfnamefont{L.}~\bibnamefont{Durlofsky}}, \bibinfo
  {author} {\bibfnamefont{J.~F.}\ \bibnamefont{Brady}},\ and\ \bibinfo {author}
  {\bibfnamefont{G.}~\bibnamefont{Bossis}},\ }%
  \bibfield{journal}{%
  \bibinfo {journal} {J. Fluid Mech.}\ }%
  \textbf{\bibinfo {volume} {180}},\ \bibinfo {pages} {21} (\bibinfo {year}
  {1987})%
  \bibAnnoteFile{NoStop}{Durlofsky87}%
\bibitem{Ermak78}%
  \BibitemOpen
  \bibfield{author}{%
  \bibinfo {author} {\bibfnamefont{D.~L.}\ \bibnamefont{Ermak}}\ and\ \bibinfo
  {author} {\bibfnamefont{J.~A.}\ \bibnamefont{McCammon}},\ }%
  \bibfield{journal}{%
  \bibinfo {journal} {J. Chem. Phys.}\ }%
  \textbf{\bibinfo {volume} {69}},\ \bibinfo {pages} {1352} (\bibinfo {year}
  {1978})%
  \bibAnnoteFile{NoStop}{Ermak78}%
\bibitem{Brady93}%
  \BibitemOpen
  \bibfield{author}{%
  \bibinfo {author} {\bibfnamefont{J.~F.}\ \bibnamefont{Brady}},\ }%
  \bibfield{journal}{%
  \bibinfo {journal} {J. Chem. Phys.}\ }%
  \textbf{\bibinfo {volume} {98}},\ \bibinfo {pages} {3335} (\bibinfo {year}
  {1993})%
  \bibAnnoteFile{NoStop}{Brady93}%
\bibitem{Kampen}%
  \BibitemOpen
  \bibfield{author}{%
  \bibinfo {author} {\bibfnamefont{N.~G.}\ \bibnamefont{van Kampen}},\ }%
  \emph{\bibinfo {title} {Stochastic Processes in Physics and Chemistry}}\
  (\bibinfo {publisher} {revised and enlarged edition, Elsevier},\ \bibinfo
  {address} {Amsterdam, The Netherlands},\ \bibinfo {year} {1992})%
  \bibAnnoteFile{NoStop}{Kampen}%
\bibitem{Landau57}%
  \BibitemOpen
  \bibfield{author}{%
  \bibinfo {author} {\bibfnamefont{L.~D.}\ \bibnamefont{Landau}}\ and\ \bibinfo
  {author} {\bibfnamefont{E.~M.}\ \bibnamefont{Lifshitz}},\ }%
  \bibfield{journal}{%
  \bibinfo {journal} {J. Exptl. Theoret. Phys. (Engl. Ed.)}\ }%
  \textbf{\bibinfo {volume} {5}},\ \bibinfo {pages} {512} (\bibinfo {year}
  {1957})%
  \bibAnnoteFile{NoStop}{Landau57}%
\bibitem{LifshitzPitaevskii}%
  \BibitemOpen
  \bibfield{author}{%
  \bibinfo {author} {\bibfnamefont{E.~M.}\ \bibnamefont{Lifshitz}}\ and\
  \bibinfo {author} {\bibfnamefont{L.~P.}\ \bibnamefont{Pitaevskii}},\ }%
  \emph{\bibinfo {title} {Statistical Physics, Part 2}},\ \bibinfo {series} {in
  Landau and Lifshitz \textit{Course of Theoretical Physics}}, Vol.~\bibinfo
  {volume} {9}\ (\bibinfo {publisher} {Butterworth and Heinemann},\ \bibinfo
  {address} {Oxford, U.K.},\ \bibinfo {year} {2004})%
  \bibAnnoteFile{NoStop}{LifshitzPitaevskii}%
\bibitem{Zwanzig64}%
  \BibitemOpen
  \bibfield{author}{%
  \bibinfo {author} {\bibfnamefont{R.}~\bibnamefont{Zwanzig}},\ }%
  \bibfield{journal}{%
  \bibinfo {journal} {J. Res. Natl. Bur. Stand. B}\ }%
  \textbf{\bibinfo {volume} {68}},\ \bibinfo {pages} {143} (\bibinfo {year}
  {1964})%
  \bibAnnoteFile{NoStop}{Zwanzig64}%
\bibitem{Fox70}%
  \BibitemOpen
  \bibfield{author}{%
  \bibinfo {author} {\bibfnamefont{R.~F.}\ \bibnamefont{Fox}}\ and\ \bibinfo
  {author} {\bibfnamefont{G.~E.}\ \bibnamefont{Uhlenbeck}},\ }%
  \bibfield{journal}{%
  \bibinfo {journal} {Phys. Fluids}\ }%
  \textbf{\bibinfo {volume} {13}},\ \bibinfo {pages} {1893} (\bibinfo {year}
  {1970})%
  \bibAnnoteFile{NoStop}{Fox70}%
\bibitem{Hauge73}%
  \BibitemOpen
  \bibfield{author}{%
  \bibinfo {author} {\bibfnamefont{E.~H.}\ \bibnamefont{Hauge}}\ and\ \bibinfo
  {author} {\bibfnamefont{A.}~\bibnamefont{Martin-L{\"o}f}},\ }%
  \bibfield{journal}{%
  \bibinfo {journal} {J. Stat. Phys.}\ }%
  \textbf{\bibinfo {volume} {7}},\ \bibinfo {pages} {259} (\bibinfo {year}
  {1973})%
  \bibAnnoteFile{NoStop}{Hauge73}%
\bibitem{Bedeaux74}%
  \BibitemOpen
  \bibfield{author}{%
  \bibinfo {author} {\bibfnamefont{D.}~\bibnamefont{Bedeaux}}\ and\ \bibinfo
  {author} {\bibfnamefont{P.}~\bibnamefont{Mazur}},\ }%
  \bibfield{journal}{%
  \bibinfo {journal} {Physica}\ }%
  \textbf{\bibinfo {volume} {76}},\ \bibinfo {pages} {247} (\bibinfo {year}
  {1974})%
  \bibAnnoteFile{NoStop}{Bedeaux74}%
\bibitem{Noetinger90}%
  \BibitemOpen
  \bibfield{author}{%
  \bibinfo {author} {\bibfnamefont{B.}~\bibnamefont{Noetinger}},\ }%
  \bibfield{journal}{%
  \bibinfo {journal} {Physica A}\ }%
  \textbf{\bibinfo {volume} {163}},\ \bibinfo {pages} {545} (\bibinfo {year}
  {1990})%
  \bibAnnoteFile{NoStop}{Noetinger90}%
\bibitem{Singh17}%
  \BibitemOpen
  \bibfield{author}{%
  \bibinfo {author} {\bibfnamefont{R.}~\bibnamefont{Singh}}\ and\ \bibinfo
  {author} {\bibfnamefont{R.}~\bibnamefont{Adhikari}},\ }%
  \bibfield{journal}{%
  \bibinfo {journal} {Euro. J. Comput. Mech.}\ }%
  \textbf{\bibinfo {volume} {26}},\ \bibinfo {pages} {78} (\bibinfo {year}
  {2017})%
  \bibAnnoteFile{NoStop}{Singh17}%
\bibitem{Bossis89}%
  \BibitemOpen
  \bibfield{author}{%
  \bibinfo {author} {\bibfnamefont{G.}~\bibnamefont{Bossis}}\ and\ \bibinfo
  {author} {\bibfnamefont{J.~F.}\ \bibnamefont{Brady}},\ }%
  \bibfield{journal}{%
  \bibinfo {journal} {J. Chem. Phys.}\ }%
  \textbf{\bibinfo {volume} {91}},\ \bibinfo {pages} {1866} (\bibinfo {year}
  {1998})%
  \bibAnnoteFile{NoStop}{Bossis89}%
\bibitem{Daivis94}%
  \BibitemOpen
  \bibfield{author}{%
  \bibinfo {author} {\bibfnamefont{P.~J.}\ \bibnamefont{Daivis}}\ and\ \bibinfo
  {author} {\bibfnamefont{D.~J.}\ \bibnamefont{Evans}},\ }%
  \bibfield{journal}{%
  \bibinfo {journal} {J. Chem. Phys.}\ }%
  \textbf{\bibinfo {volume} {100}},\ \bibinfo {pages} {541} (\bibinfo {year}
  {1994})%
  \bibAnnoteFile{NoStop}{Daivis94}%
\bibitem{Batchelor70}%
  \BibitemOpen
  \bibfield{author}{%
  \bibinfo {author} {\bibfnamefont{G.~K.}\ \bibnamefont{Batchelor}},\ }%
  \bibfield{journal}{%
  \bibinfo {journal} {J. Fluid Mech.}\ }%
  \textbf{\bibinfo {volume} {41}},\ \bibinfo {pages} {545} (\bibinfo {year}
  {1970})%
  \bibAnnoteFile{NoStop}{Batchelor70}%
\bibitem{Palanisamy19}%
  \BibitemOpen
  \bibfield{author}{%
  \bibinfo {author} {\bibfnamefont{D.}~\bibnamefont{Palanisamy}}\ and\ \bibinfo
  {author} {\bibfnamefont{W.~K.}\ \bibnamefont{den Otter}},\ }%
  \bibinfo {note} {in preparation}%
  \bibAnnoteFile{NoStop}{Palanisamy19}%
\bibitem{Ottinger}%
  \BibitemOpen
  \bibfield{author}{%
  \bibinfo {author} {\bibfnamefont{H.~C.}\ \bibnamefont{{\"O}ttinger}},\ }%
  \emph{\bibinfo {title} {Stochastic Processes in Polymeric Fluids}}\ (\bibinfo
  {publisher} {Springer-Verlag},\ \bibinfo {address} {Berlin, Germany},\
  \bibinfo {year} {1996})%
  \bibAnnoteFile{NoStop}{Ottinger}%
\bibitem{Gardiner}%
  \BibitemOpen
  \bibfield{author}{%
  \bibinfo {author} {\bibfnamefont{C.}~\bibnamefont{Gardiner}},\ }%
  \emph{\bibinfo {title} {Stochastic Methods. A handbook for the Natural and
  Social Sciences}},\ \bibinfo {edition} {4th}\ ed.,\ Springer Series in
  Synergetics\ (\bibinfo {publisher} {Springer-Verlag},\ \bibinfo {address}
  {Berlin, Germany},\ \bibinfo {year} {2009})%
  \bibAnnoteFile{NoStop}{Gardiner}%
\bibitem{Einstein05}%
  \BibitemOpen
  \bibfield{author}{%
  \bibinfo {author} {\bibfnamefont{A.}~\bibnamefont{Einstein}},\ }%
  \bibfield{journal}{%
  \bibinfo {journal} {Ann. Phys.}\ }%
  \textbf{\bibinfo {volume} {17}},\ \bibinfo {pages} {549} (\bibinfo {year}
  {1905})%
  \bibAnnoteFile{NoStop}{Einstein05}%
\bibitem{Batchelor76}%
  \BibitemOpen
  \bibfield{author}{%
  \bibinfo {author} {\bibfnamefont{G.~K.}\ \bibnamefont{Batchelor}},\ }%
  \bibfield{journal}{%
  \bibinfo {journal} {J. Fluid Mech.}\ }%
  \textbf{\bibinfo {volume} {74}},\ \bibinfo {pages} {1} (\bibinfo {year}
  {1976})%
  \bibAnnoteFile{NoStop}{Batchelor76}%
\bibitem{Palanisamy18}%
  \BibitemOpen
  \bibfield{author}{%
  \bibinfo {author} {\bibfnamefont{D.}~\bibnamefont{Palanisamy}}\ and\ \bibinfo
  {author} {\bibfnamefont{W.~K.}\ \bibnamefont{den Otter}},\ }%
  \bibfield{journal}{%
  \bibinfo {journal} {J. Chem. Phys.}\ }%
  \textbf{\bibinfo {volume} {148}},\ \bibinfo {pages} {194112} (\bibinfo {year}
  {2018})%
  \bibAnnoteFile{NoStop}{Palanisamy18}%
\bibitem{Makino04motion}%
  \BibitemOpen
  \bibfield{author}{%
  \bibinfo {author} {\bibfnamefont{M.}~\bibnamefont{Makino}}\ and\ \bibinfo
  {author} {\bibfnamefont{M.}~\bibnamefont{Doi}},\ }%
  \bibfield{journal}{%
  \bibinfo {journal} {J. Phys. Soc. Japan}\ }%
  \textbf{\bibinfo {volume} {73}},\ \bibinfo {pages} {2739} (\bibinfo {year}
  {2004})%
  \bibAnnoteFile{NoStop}{Makino04motion}%
\bibitem{Ilie15}%
  \BibitemOpen
  \bibfield{author}{%
  \bibinfo {author} {\bibfnamefont{I.~M.}\ \bibnamefont{Ilie}}, \bibinfo
  {author} {\bibfnamefont{W.~J.}\ \bibnamefont{Briels}},\ and\ \bibinfo
  {author} {\bibfnamefont{W.~K.}\ \bibnamefont{den Otter}},\ }%
  \bibfield{journal}{%
  \bibinfo {journal} {J. Chem. Phys.}\ }%
  \textbf{\bibinfo {volume} {142}},\ \bibinfo {pages} {114103} (\bibinfo {year}
  {2015})%
  \bibAnnoteFile{NoStop}{Ilie15}%
\bibitem{Rallison78}%
  \BibitemOpen
  \bibfield{author}{%
  \bibinfo {author} {\bibfnamefont{J.~M.}\ \bibnamefont{Rallison}},\ }%
  \bibfield{journal}{%
  \bibinfo {journal} {J. Fluid Mech.}\ }%
  \textbf{\bibinfo {volume} {84}},\ \bibinfo {pages} {237} (\bibinfo {year}
  {1978})%
  \bibAnnoteFile{NoStop}{Rallison78}%
\bibitem{Makino04visco}%
  \BibitemOpen
  \bibfield{author}{%
  \bibinfo {author} {\bibfnamefont{M.}~\bibnamefont{Makino}}\ and\ \bibinfo
  {author} {\bibfnamefont{M.}~\bibnamefont{Doi}},\ }%
  \bibfield{journal}{%
  \bibinfo {journal} {J. Phys. Soc. Japan}\ }%
  \textbf{\bibinfo {volume} {73}},\ \bibinfo {pages} {3020} (\bibinfo {year}
  {2004})%
  \bibAnnoteFile{NoStop}{Makino04visco}%
\end{thebibliography}

%

\end{document}